\documentstyle[preprint,aps,prl,psfig]{revtex}

\begin{document}
\topmargin -0.05cm
\title{Energetic Components of Cooperative Protein Folding }
\author{H\"useyin KAYA  and Hue Sun CHAN}
\address{Department of Biochemistry, and \\
Department of Medical Genetics \& Microbiology \\
Faculty of Medicine, University of Toronto \\
Toronto, Ontario M5S 1A8, Canada}

\def\kB{{k_{\scriptscriptstyle B}}}
\def\gD{{g_{\scriptscriptstyle{\rm D}}}}
\def\gcD{{{\rm g}_{\scriptscriptstyle{\rm D}}}}
\def\HD{{H_{\rm D}}}
\def\sH{{\sigma_H}}
\def\eHDkT{{{\rm e}^{-H_{\rm D}/(\kB T)}}}
\def\eHkT{{{\rm e}^{-H/(\kB T)}}}
\def\DHv{{\Delta H_{\rm vH}}}
\def\DHc{{\Delta H_{\rm cal}}}
\def\DHvDHc{{\Delta H_{\rm vH}/\Delta H_{\rm cal}}}

\maketitle

\begin{abstract}
A new lattice protein model with a four-helix bundle ground
state is analyzed by a parameter-space Monte Carlo histogram
technique to evaluate the effects of an extensive variety of
model potentials on folding thermodynamics. Cooperative
helical formation and contact energies based on a 5-letter
alphabet are found to be insufficient to satisfy calorimetric
and other experimental criteria for two-state folding.
Such proteinlike behaviors are predicted, however, by models
with polypeptide-like local conformational restrictions and
environment-dependent hydrogen bonding-like interactions.

\end{abstract}
\eject


Proteins are complex systems. Insight into their behaviors has
been gained by simplified models of generic proteins [1-6]. 
To serve as stepping stones towards an elucidation of the physics of
real proteins, however, these models must be subjected to rigorous
evaluations against experiments. Recently, we found that  
a number of popular lattice protein models with 2, 3, and 20 residue
types and pairwise additive contact energies do not   
satisfy the experimental criteria for two-state thermodynamic
cooperativity
which, among other conditions, requires a protein's van't Hoff to
calorimetric
enthalpy ratio $\Delta H_{\rm vH}/\Delta H_{\rm cal}\approx 1$ [7,8].
While certain G\=o models are cooperative,
they do not address the physical nature of protein interactions since
their potentials are {\it teleological} [8] in the sense that
the energetic favorability of the native conformation as a whole is 
presupposed.

It has since been proposed that a cooperative interplay between local
conformational preferences and nonlocal interactions could give rise to
proteinlike thermodynamics [7].
To evaluate the viability of this scenario, we introduce in this Letter a
55-mer chain model with a four-helix bundle ground state (Fig.~1) that
shares common features with the corresponding protein motif [9].
The cubic-lattice helices are right-handed
(called type (i) in Ref.~[2]), as are most $\alpha$-helices. The model
has the following energetic components. (In the analysis below all
energies are dimensionless and temperature independent.)

{\it Contact energies.}
A reduced 5-letter alphabet (Fig.~1) identical to the one
recently optimized [5] is used for nearest-neighbor interactions,   
with energy parameters from Table~III of Kolinski et al. [10].
While these energies are proteinlike to some extent because
hydrophobic groups are placed in the native core (Fig.~1), they do not
represent the full interactions between amino acid residues [10].
Here they are adopted to capture
heterogeneous aspects of intraprotein interactions [11].

{\it Unfavorable local conformations.}
Two types of local non-proteinlike bond geometries
are discouraged: the initiation of a left-handed (lh) helix (Fig.~2a),
and one end of a helix taking a sharp turn (st) to fold back
onto itself (Fig.~2b) are assigned unfavorable ($>0$) energies
to take into account that in real
proteins left-handed $\alpha$-helices are sterically disfavored, and that
polypeptides are stiffer than a fully flexible chain [12].

{\it Environment-dependent hydrogen bonding.}
The favorable many-body interactions (${\cal E}_{\rm Hb}<0$) in Fig.~2c,d
are introduced to explore an idea, suggested by experiments [13], that the
collective strength of hydrogen bonds is stronger when they are buried in
the core of a protein than when they are exposed to water.
Hence we focus mainly on $b>1$ cases below. Analogous   
interactions have been used before [7,14].

{\it Cooperative helical propagation.}
An extra favorable energy is assigned to every two consecutive helical
turns (Fig.~2e) to encourage helix elongation. Such an effect
may arise from dipole-dipole interactions between amide groups
in real $\alpha$-helices [12].   

The total energy of a conformation from these contributions is
$$
E= E_{\rm contact} + \gamma_{\rm lh}N_{\rm lh}
+ \gamma_{\rm st}N_{\rm st} + {\cal E}_{\rm Hb} N^{(6)}_{\rm Hb}
+ b{\cal E}_{\rm Hb} N^{(8)}_{\rm Hb} + {\cal E}_{\rm Helix}N_{\rm Helix} 
\; ,
\eqno(1)
$$
where $E_{\rm contact}$ is the sum of 5-letter
contact energies, $N_{\rm lh}$ and $N_{\rm st}$
are respectively the numbers of all incidences of Fig.~2a,b.
In the present analysis, $N_{\rm Helix}$ only counts those helices
(Fig.~2e) that are parts of the four helices in the ground-state
conformations (monomers 1--12, 15--26, 30--41, and 44--55; see Fig.~1);
and the hydrogen bonding pairs counted by $N^{(6)}_{\rm Hb}$
(Fig.~2c) and $N^{(8)}_{\rm Hb}$ (Fig.~2d) are the 36 $(i,i+3)$ or
$(i,i+5)$ contacting monomer pairs in the native helices.
A first-principle treatment would have assigned
hydrogen bonds and helical segments in a manner that do not require
knowledge of the native structure. However, progress can nonetheless
be made by the approach taken here,
which seeks, as a first step in the inquiry, to ascertain
the consequence of presupposing {\it local}
native preferences, a presupposition that is notably less dependent on
{\it
a priori} knowledge than the assumption of {\it global} native preference
in G\=o potentials.

To efficiently determine how thermodynamic properties vary with the model
energetic parameters, we use a generalization [3,15]
of the standard Metropolis Monte Carlo histogram technique [8,16]
to eliminate the need to perform separate direct simulations
for every parameter set of interest.
Typical simulations are carried out with ${\cal E}_{\rm Hb}$ $=$
${\cal E}_{\rm Helix}=0$ at a certain temperature $T^\prime$, during which
numbers $P$ of sampled conformations are binned into a
multiple-dimensional
array (histogram) according to
$(E^\prime,N^{(6)}_{\rm Hb},N^{(8)}_{\rm Hb},N_{\rm Helix})$, where
$E^\prime$ is the energy of the conformation. Thus the density
of states of the simulated system
$g(E^\prime,N^{(6)}_{\rm Hb},N^{(8)}_{\rm Hb},N_{\rm Helix})$
$=$ $P(E^\prime,N^{(6)}_{\rm Hb},N^{(8)}_{\rm Hb},N_{\rm Helix})
e^{E^\prime/\kB T^\prime}$, where $\kB$ is Boltzmann's constant.
It follows that the partition function for any ${\cal E}_{\rm Hb}$,
${\cal E}_{\rm Helix}$ at any temperature $T$ is given by $\sum
g(E^\prime,N^{(6)}_{\rm Hb},N^{(8)}_{\rm Hb},N_{\rm Helix})e^{-E/\kB T}$,
where the summation is over
$E^\prime$, $N^{(6)}_{\rm Hb}$, $N^{(8)}_{\rm Hb}$, and $N_{\rm Helix}$,
and
$E=E^\prime + {\cal E}_{\rm Hb}(N^{(6)}_{\rm Hb}
+ bN^{(8)}_{\rm Hb}) + {\cal E}_{\rm Helix}N_{\rm Helix}$.
A similar procedure is used to study the effects of
$\gamma_{\rm lh}$ and $\gamma_{\rm st}$.

Each Monte Carlo run consists of $1.53\times 10^9$ attempted
moves, the first $3\times 10^7$ of which are
excluded from data acquisition. To generate a multiple-dimensional
histogram, a total of 20 runs at 10 different simulation $T^\prime$s
(around the transition region) are performed, with two different random
initial conformations for each $T^\prime$. Values of $\kappa_2$ estimated
from different $T^\prime$s agree well, with
standard deviation $\approx 4\%$. We have conducted extensive comparisons
with direct simulations to validate the method [17].
No energy lower than that of the structure in Fig.~1 has been encountered.

As in experimental calorimetry [18], we characterize the
thermodynamic cooperativity of a model protein by its specific heat
capacity
$C_P$ and $\Delta H_{\rm vH}/\Delta H_{\rm cal}$.  
When baseline subtraction is not applied to the $C_P$ function,
$\Delta H_{\rm vH}/\Delta H_{\rm cal}$ may be equated to
$\kappa_2\equiv$ $2T_{\rm max}\sqrt{\kB C_P(T_{\rm max})}/\Delta H_{\rm
cal}$,
where $\Delta H_{\rm cal}=\int_0^\infty dT C_P(T)$ is the calorimetric
enthalpy, and $C_P(T)$ is maximum at $T=T_{\rm max}$ [8].
Baseline subtractions amount to
defining a multi-conformation native state
and ignoring a part of the enthalpic variation in the denatured ensemble.
This effectively reduces both the calorimetric enthalpy and the maximum
heat
capacity value, resulting in a modified enthalpy ratio
$\kappa_2^{({\rm s})}$ ($>\kappa_2$) [8].

Our main findings are summarized in Figs.~3 and 4.
The apparent $\Delta H_{\rm vH}/\Delta H_{\rm cal}$ $=\kappa_2^{({\rm
s})}$
after empirical baseline subtractions can often be close or equal to unity  
even when $\kappa_2$ is low (Fig.~4).
However, as is the case for a 3-letter model, a large discrepancy
between $\kappa_2$ and $\kappa_2^{({\rm s})}$ is often symptomatic of
non-proteinlike significant post-denaturational chain expansion at
$T\gg T_{\rm max}$ [8]. Therefore, for model evaluation, proteinlike  
thermodynamic cooperativity requires {\it both} a small  
$\kappa_2^{({\rm s})}$ $-$ $\kappa_2$ {\it and}
$\kappa_2^{({\rm s})}\approx 1$.

Fig. 3 compares three models by this criterion. The least cooperative
is a flexible chain model with only pairwise additive contact energies.
A second model with polypeptide-like local sterics has
slightly enhanced cooperativity because populations of nonnative
conformations with
non-proteinlike local geometries that are hitherto favorable in a fully
flexible chain model are reduced. However, both of these models
are not proteinlike because of their significant post-denaturational chain
expansions, as is evident from their thick denatured $C_P$ ``tails'' at
$T\gg T_{\rm max}$ (Fig.~3), which account for these models' relatively
large differences between 
$\kappa_2^{({\rm s})}$ and $\kappa_2$ [8]. On the other hand, the
model that also incorporates environment-dependent hydrogen
bonding has more proteinlike thermodynamics:
Its native $C_P$ tail at $T\ll T_{\rm max}$
is thin, with $C_P$ values lower on average than that
of its denatured tail (Fig.~3).
This implies that its native conformational diversity
is limited, thus conforming better to NMR
data [19] than a previously considered 20-letter model [8].
As for real proteins [20], its average radius of gyration undergoes a
sharp change around $T_{\rm max}$, but has no appreciable
post-denaturational
increase (data not shown).

Fig.~4 surveys a range of energetic parameters.
Remarkably, local helical cooperativity has only a small
impact on overall folding cooperativity, and proteinlike
thermodynamics is possible at ${\cal E}_{\rm Helix}=0$.
While ${\cal E}_{\rm Helix}<0$ stabilizes the native state, it also
stabilizes
denatured conformations with partially intact native helices.
Therefore, its effect on calorimetric cooperativity is not substantial
because it cannot widen the average enthalpy difference between native
and denatured states significantly [7,8]. In contrast, calorimetric
cooperativity increases sharply with more negative ${\cal E}_{\rm Hb}$.
For $0.5\le b\le 1.5$, this effect is not very sensitive to
$b$. For example, ${\cal E}_{\rm Hb}=-0.5$, ${\cal E}_{\rm Helix}=0$,
$b=0.5$ (and $\gamma_{\rm lh}=6.0$, $\gamma_{\rm st}=5.0$) lead to   
$\kappa_2$, $\kappa_2^{({\rm s})}=$ $0.85$, $0.99$, which are only
slightly lower than the $\kappa_2$, $\kappa_2^{({\rm s})}$
values of $0.90$, $1.0$ for the same ${\cal E}_{\rm Hb}$ and
${\cal E}_{\rm Helix}$ in Fig.~4 for $b=1.5$.  

In addition, we found that a lattice version of
helix capping [21] has a slight attenuating effect on cooperativity [17].
Consistent with experiments [22], in our model, tertiary interactions
are essential in stablizing helices in native structures.
When folded independently, the two 12-mer sequences
for the native helices at 1--12 and 15--26 are much less stable
($T_{\rm max}\sim 0.3$), and their thermal transitions are not
calorimetrically cooperative.

Figs.~3 and 4 suggest that a cooperative interplay between local
conformational preferences and nonlocal contact interactions, as
exemplified
by the environment-dependent hydrogen bonding in the model, is a viable
mechanism for proteinlike thermodynamics;
and that the required cooperative effect ($b{\cal E}_{\rm Hb}=-0.8$)
need not be exceedingly strong relative to the pairwise contact energies
(average magnitude $=0.66$). This observation is consistent with a
previous
high-coordination lattice model study [23], though the latter did not
address the calorimetric criterion. 
The present approach did not consider non-native
hydrogen bonding. Cooperativity in the present model would be
reduced if such non-native conformations are favored.
Further investigations using continuum models with
polypeptide chain geometry are necessary to ascertain  
whether real proteins can have substantial number of
non-native hydrogen bonds.
While the present model
should be regarded as tentative because it
relies on local native information, its proteinlike features
do not follow trivially from part of its interactions' native-centric
nature {\it per se}.
Important physical principles have emerged from our analysis  
because not all native-centric interaction schemes can bring about
comparable enhancements in thermodynamic cooperativity: ({\it i}) The fact
that
${\cal E}_{\rm Helix}<0$ is neither necessary nor sufficient
for proteinlike thermodynamics suggests that multi-body interactions that   
favors local native conformation irrespective of tertiary packing cannot
account for calorimetric cooperativity.  
({\it ii}) A G\=o model for the ground-state conformations in Fig.~1,
which exploits both local and nonlocal native information but is based  
exclusively on pairwise additive interactions, has $\kappa_2=0.73$ and is  
thus less cooperative than the model with $\kappa_2=0.91$ in Fig.~3. 
({\it iii}) Proteinlike steric effects contribute
to cooperativity. If non-proteinlike local
conformations were not disfavored in the latter model
(i.e., if $\gamma_{\rm lh}=\gamma_{\rm st}=0$),
$\kappa_2$ would be reduced to $0.85$.

We have thus mapped out a general investigative strategy and established
the
viability of a folding scenario. It should be emphasized, however,
that satisfying the
requirements for thermodynamic cooperativity is clearly necessary but not    
sufficient for the validity of a scenario's underlying physical
mechanisms.
Whether hydrogen bonding is favorable to native stability remains
controversial [13,24]. The present choice of ${\cal E}_{\rm Hb}<0$   
is motivated by experiments [13]. But there have been theoretical
suggestions
that hydrogen bonding disfavors the folded state of a protein [24].
In the present modeling framework, that would be detrimental to
calorimetric
cooperativity as it corresponds to ${\cal E}_{\rm Hb}>0$, leading to
$\kappa_2$s even smaller than the ${\cal E}_{\rm Hb}=0$ case (Fig.~3).
For instance, when ${\cal E}_{\rm Hb}=+0.1$ and ${\cal E}_{\rm Helix}=0$,
(and $\gamma_{\rm lh}=6.0$, $\gamma_{\rm st}=5.0$),
$\kappa_2$, $\kappa_2^{({\rm s})}=$ $0.51$, $0.80$ for $b=0.5$ and
$\kappa_2$, $\kappa_2^{({\rm s})}=$
$0.45$, $0.70$ for $b=1.5$, respectively. If that turns out to be the
case, 
there would be added impetus to extend the present method
to ascertain the role of other mechanisms such as sidechain packing
[4,8,25] in
protein calorimetric two-state cooperativity.


This work was supported by Medical Research Council of Canada
grant MT-15323.


\par\vfill\eject





\par\vfill\eject


\noindent
{\large\bf Figure Captions}\\

{\bf Fig.~1}. Ground state conformations of the model.
Black beads denote nominally hydrophobic monomers.
The numbers label selected sequence positions.
Each of the 3 short loops
at positions (13, 14), (27, 28, 29), and (42, 43) has two
iso-energetic local conformations. Thus the ground state has 8
conformations, one of which is shown.
\\

{\bf Fig.~2}. Energetic components of the model.
(a,b) Each contact marked by a double arrow is assigned an energy (as
shown).
The dotted lines in (b) depict an alternate path of the chain from  
monomer $i-3$ to $i$ that has one instead of two unfavorable contacts.
Similarly, an energy is associated with each buried hydrogen bond (c,d)
and
each occurrence of two consecutive turns (layers) of a right-handed
lattice
helix (e).  In (c,d), hydrogen bonds are represented as ladders linking
pairs
of encircled monomers. Their burial requires occupation of at least 6 of
their
neighbor sites. The energy of a completely buried bond with 8 occupied
neighbor sites (d) can be stronger ($b>1$) or weaker ($b<1$) than
a partially buried bond with 6 or 7 occupied neighbor sites. The
latter two cases have the same energy ${\cal E}_{\rm Hb}$
and are accounted for collectively by $N^{(6)}_{\rm Hb}$ in Eq.~(1).
\\
  
{\bf Fig.~3}. Specific heat capacity functions.
In all three cases shown, ${\cal E}_{\rm Helix}=0$.
$T_{\rm max}$s are marked by vertical lines.
>From left to right, the first model has only the 5-letter pairwise
contact
energies ($\gamma_{\rm lh}=\gamma_{\rm st}={\cal E}_{\rm Hb}=0$).
In addition to these, the second model incorporates the
repulsive interactions in Fig.~2a,b, with $\gamma_{\rm lh}=6.0$ and
$\gamma_{\rm st}=5.0$. The third model further adds
the favorable hydrogen-bonding
energies in Fig.~2c,d, with ${\cal E}_{\rm Hb}=-0.53$,
$b{\cal E}_{\rm Hb}=-0.8$, and $b=1.5$.
$\kappa_2=0.55$, $0.62$, and $0.91$, and
after the plotted inclined baselines are subtracted,
$\kappa_2^{({\rm s})}=0.93$, $0.93$, and $1.01$, respectively.
\\
  
{\bf Fig.~4}.
Calorimetric cooperativity as a function of local helical cooperativity
and environment-dependent hydrogen bonding strength, with $b=1.5$,
$\gamma_{\rm lh}=6.0$ and $\gamma_{\rm st}=5.0$.
$\kappa_2^{({\rm s})}$ and $\kappa_2$ are
given by the upper and lower surfaces, respectively.   
The black dots connected by a vertical bar mark the
parameters for the most cooperative case in Fig.~3.
$\kappa_2^{({\rm s})}$ are calculated using empirical
baselines [8] constructed as tangents of the $C_P$ function at
$C_P^{\prime\prime}(T)/ C_P^{\prime\prime}(T_{\rm max})=-0.001$,
where $C_P^{\prime\prime}\equiv d^2 C_P/dT^2$ (see Fig.~3).
\\

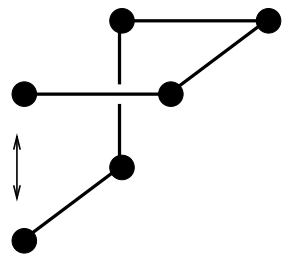
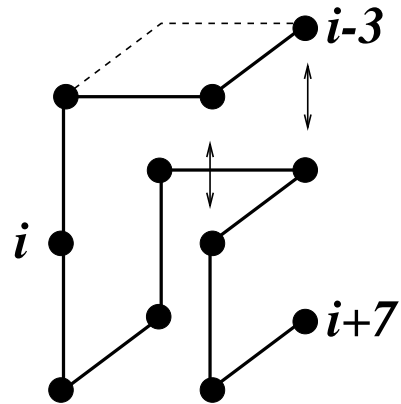
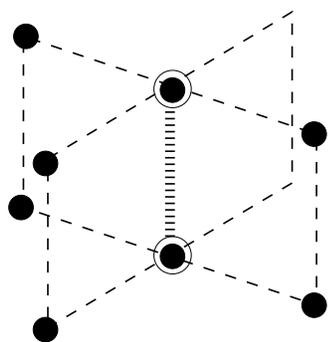
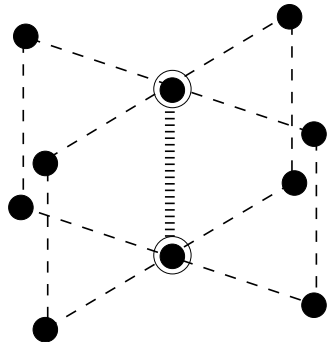
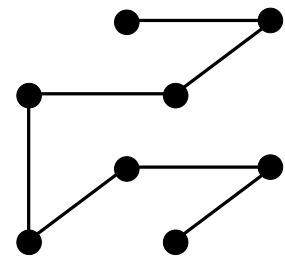
\end{document}